\documentclass[proof,mathleft]{WileyASNA-v1}

\articletype{Original Article}
\received{4 June 2024}
\revised{4 July 2024}
\accepted{4 August 2024}

\usepackage{graphicx}
\usepackage{times}
\usepackage{hyperref}
\usepackage{amsmath}
\usepackage{multicol,array}

\hypersetup{
  hidelinks,
  colorlinks=false,
}

\usepackage{times}

\usepackage{natbib}
\bibpunct{(}{)}{;}{a}{}{,}

\usepackage{color,ulem,soul}

\definecolor{dark-red}{rgb}{0.75, 0.00, 0.00}
\definecolor{hlcolor}{rgb}{1.00, 0.94, 0.92}\sethlcolor{hlcolor}

\usepackage{xspace}
\newcommand{\nirv}{\textsc{nirvana-iii}\xspace}

\newcommand{\Fig}[1]{Fig.~\ref{fig:#1}{\hspace{-1ex}}}
\newcommand{\Figure}[1]{Figure~\ref{fig:#1}{\hspace{-1ex}}}

\begin{document}

\title{Toward an efficient second-order method for computing the surface gravitational potential on spherical-polar meshes}

\author[1]{Oliver Gressel}
\author[1]{Udo Ziegler}

\authormark{Gressel \& Ziegler}

\address[1]{
  \orgdiv{$\!$MHD \& Turbulence Section},\newline
  \orgname{Leibniz Institute for Astrophysics Potsdam},
  \orgaddress{An der Sternwarte 16,\newline 14482 Potsdam,\country{Germany}}
}

\corres{*\,\email{ogressel@aip.de}}


\abstract{Astrophysical accretion discs that carry a significant mass compared with their central object are subject to the effect of self-gravity. In the context of circumstellar discs, this can, for instance, cause fragmentation of the disc gas, and --under suitable conditions-- lead to the direct formation of gas-giant planets. If one wants to study these phenomena, the disc's gravitational potential needs to be obtained by solving the Poisson equation. This requires to specify suitable boundary conditions. In the case of a spherical-polar computational mesh, a standard multipole expansion for obtaining boundary values is not practicable. We hence compare two alternative methods for overcoming this limitation. The first method is based on a known Green's function expansion (termed ``CCGF'') of the potential, while the second (termed ``James' method'') uses a surface screening mass approach with a suitable discrete Green's function. We demonstrate second-order convergence for both methods and test the weak scaling behaviour when using thousands of computational cores. Overall, James' method is found superior owing to its favourable algorithmic complexity of $\sim \mathcal{O}(n^3)$ compared with the $\sim\mathcal{O}(n^4)$ scaling of the CCGF method.}

\keywords{Accretion discs --- self gravity --- gravitational instability --- Poisson solver}

\jnlcitation{\cname{    \author{O.~Gressel}
    \author{U.~Ziegler}
  } (\cyear{2024}),
  \ctitle{Toward an efficient second-order method for computing the surface gravitational potential on spherical-polar meshes},
  \cjournal{Astronomical Notes}, \cvol{2024;00:x--y}.
}

\fundingInfo{Funded/Co-funded by the European Union (via ERC Consolidator Grant, \textsc{epoch-of-taurus}, project number 101043302). Views and opinions expressed are however those of the author(s) only and do not necessarily reflect those of the European Union or the European Research Council. Neither the European Union nor the granting authority can be held responsible for them.}

\maketitle


\def\nat{Nature\ }
\def\aa{ Astronomy \& Astrophysics\ }
\def\aap{ Astronomy \& Astrophysics\ }
\def\aas{ Astronomy \& Astrophysics Suppl.}
\def\ara\&a{ Ann. Rev. Astronomy Astrophysics}
\def\aarev{ Astronomy \& Astrophysics Rev.}
\def\apj{ The Astrophysical Journal\ }
\def\apjl{ The Astrophysical Journal Letters\ }
\def\apjs{ Astrophys. J. Suppl.}
\def\apss{ Astrophys. Space Sci.}
\def\aj{ Astronomical Journal}
\def\jfm{ J. Fluid Mechanics\ }
\def\arafm{Ann. Rev. Fluid Mechanics\ }
\def\gafd{Geophysical Astrophysical Fluid Dyn.\ }
\def\mnras{Month. Not. Roy. Astr. Soc.\ }
\def\an{ Astronomische Nachrichten\ }
\def\sp{ Solar Physics\ }
\def\grl{ Geophysical Research Letters\ }
\def\jgr{ J. Geophysical Research\ }
\def\prl{Physical Review Letters\ }
\def\pre{ Physical Review E\ }
\def\prb{ Physical Review B\ }
\def\ssr{ Space Science Review\ }


\section{Introduction} \label{sec:intro}

There exist many astrophysical contexts, where the collapse and fragmentation of a gaseous object subject to its own gravity plays a key role -- such as, for instance, the collapse of a molecular cloud in the interstellar medium, leading to the formation of stars and eventually planets \citep{2007ARA&A..45..565M}. Often times, owing to the initial rotation of the gas cloud, a rotationally supported disc is formed in the process and acts as an intermediate conduit for the accreting material \citep{2023ASPC..534..317T}. Depending on the relative importance of disc surface gravity (versus the gravity of the central object), on the one hand, and the ability of the gas to cool efficiently, on the other hand, the disc itself may be subject to gravitational fragmentation. In its non-linear stage, the instability can lead to the formation of spiral features and gravitationally-bound clumps \citep[see, e.g.][for a recent study on possible morphologies]{2021A&A...650A..49B}. Under suitable conditions (found preferentially at large orbital radii), it is feasible to directly form giant planets or brown dwarfs via this pathway \citep[see][]{2022oeps.book..250R}.

To study the evolution of a rotating, self-gravitating disc accurately, when employing a Eulerian hydrodynamics code, it is often advisable to adopt a computational mesh in a polar coordinate system. For the situation in which the orientation of the rotation axis is known (and fixed in time), such a polar description can be shown to conserve angular momentum to machine accuracy. This is opposed to the situation encountered with a Eulerian description based on Cartesian coordinates, where only the {\it linear} momenta along the coordinate axes are conserved quantities by construction, and angular flows are in general rather poorly preserved \citep{2023A&A...678A.134J}.

Including the origin and/or axis of the polar grid is often discouraged, since this introduces singularities that need to be regularised. Moreover, the central grid anisotropy poses severe limitations with respect to the computational timestep owing to the shrinking separations inherent in the polar grid. For this reason, various de-refinement strategies have been proposed \citep[e.g.][]{2018ApJ...857...34Z,2021ApJ...907...43H,2023A&A...677A...9L}, though mostly for the case without considering gas self-gravity. On the other hand, adopting a polar grid comes at the price of introducing a central hole in the overall mesh topology.

This brings us to an important aspect that we need to address if we want to solve the Poisson equation for self-gravity, namely providing consistent boundary conditions for the gravitational potential on the surface of the computational domain \citep[see, e.g., the discussion in][]{2005A&A...435..385Z}. In practice, a Dirichlet-type condition demands specifying the value of the potential on the boundary. For the case of a suitably confined matter distribution, a feasible approximation is usually obtained by pushing the boundaries sufficiently far away and then evaluating a multipole expansion of the potential field. In the case of the central inner hole in the mesh topology that we encounter with polar meshes, such an approach is, however, doomed to fail, since the underlying geometrical assumption of the multipole expansion is not satisfied.

While it is ---at least in principle--- possible to evaluate the surface potential in the form of a three-dimensional convolution integral, the scaling with the number of resolution elements is often computationally prohibitive. This is because each space dimension needs to be accounted for twice in the Green's function (GF) kernel -- once at each location where the potential is evaluated, and once at each density source point. For instance, if we need to calculate $\Phi(r=r_{\rm surf},\theta=\theta_{\rm surf},\phi)$ along the one-dimensional curve comprising the poloidal boundary, and moreover for each azimuthal location, $\phi$, this adds two dimensions to the convolution operation. As we have to take into account the density $\rho(r,\theta,\phi)$ in the entire bulk volume, this adds three more dimensions. In total, this already amounts to a complexity order of $\mathcal{O}(n^5)$. One possible way of addressing the issue is to employ a ``compact cylindrical'' Green's function expansion \citep[][termed `CCGF' in the following]{1999ApJ...527...86C}, utilising elliptic integral functions. Here, employing a spectral decomposition in azimuth, one can break down the $\mathcal{O}(n^2)$ complexity associated with the $(\phi,\phi')$ coordinate pair in the GF to $\mathcal{O}(n\,\log(n))$ -- rendering the method $\approx\mathcal{O}(n^4)$ overall. Note that this only becomes possible by exploiting the favourable scaling of the fast Fourier transform (FFT) algorithm. Despite this success, further restrictions need to be kept in mind that arise from the reduced efficiency of the FFT algorithm in the framework of distributed-memory parallelism via the message passing interface (MPI). In principle, the expansion in the azimuthal wavenumber, $m$, used in the CCGF method can be artificially truncated at some arbitrary $m=m_{\rm max}$ for the sake of reduced computational expense. In terms of the accuracy of the method, such an approach may not always be justified \citep[see, e.g.,][]{2004ApJ...616..364F}.

To overcome the debilitating limitations described above, \citet{2019ApJS..241...24M} have recently presented an alternative second-order scheme for efficiently solving the Poisson equation for self-gravity on cylindrical meshes. Their approach re-introduces the classic scheme of \citet{1977JCoPh..25...71J}, and promises an overall complexity of $\approx\mathcal{O}(n^3)$.

For discs of constant thickness (as a function of radius), a cylindrical grid is typically a reasonable choice. In contrast, for the case of a moderately flaring disc --- i.e., a disc with an approximately constant aspect ratio of the disc thickness versus the distance from the star --- a spherical-polar mesh is often adopted instead. In this case, the coordinate describing the latitude is nicely adapted to the disc geometry, distributing resolution elements homogeneously, i.e., independently of the radial location. Enabling to exploit this advantage is our main motivation for extending the work of \citet{2019ApJS..241...24M} to the case of spherical-polar mesh geometry in the presented article.


The paper is organised as follows: In Section~\ref{sec:methods}, we describe our novel numerical methodology, and present a convergence study in the following Section~\ref{sec:conv}. We demonstrate the computational performance of our implementation in Section~\ref{sec:bench}, and draw our conclusions in Section~\ref{sec:conclusions}. Finally, in the appendix, we provide a closed-form analytic solution for the gravitational potential of a homogeneous spherical-polar mesh segment.


\section{Methods} \label{sec:methods}

For solving the three-dimensional Poisson equation,
\begin{equation}
  \nabla^2 \Phi = 4\pi G\, \rho\,,
  \label{eq:poisson}
\end{equation}
in the bulk volume, the adaptive-mesh \nirv code implements an efficient multigrid (MG) solver -- that is, for sufficiently isotropic grids. The method is described in detail in \citet{2005A&A...435..385Z}. Compared to the algorithm described in that paper, there is an important improvement that has been added afterwards: For the case of a uniform mesh, the algorithm is now reduced to a two-grid solver, working on the base mesh, as well as a ``sub mesh'' with a coarsened grid (by a factor of two in each space dimension). This is in contrast to the original method, which did not feature any additional mesh below the base mesh, and which reduced to a simple successive over-relaxation (SOR) scheme for non-adaptive grids.

Our approach for obtaining the values of the gravitational potential required for imposing a Dirichlet boundary condition at the poloidal domain surfaces closely follows the one take by \citet{2019ApJS..241...24M}, which is based on the original idea of \citet{1977JCoPh..25...71J}. We here briefly outline the required steps.


\subsection{The screening-mass approach} \label{sec:screen}


The basic idea, popularised by \citet{1977JCoPh..25...71J}, is derived from the fundamental equivalence of the self-gravity equation with electrostatics. In the latter realm, consider a charge distribution within a box made of conducting walls. If the encasement is electrically grounded, the electrostatic field of the charge distribution will induce surface ``screening charges'', establishing a zero electrostatic potential on the wall as well as the shielded exterior. The thus achieved solution is a superposition of two potentials, one from the actual charge distribution within the box, and one from the induced surface charges. Replacing the electric charge distribution with the mass density distribution, and terming the two potential fields and their sum as, $\Phi$, $\Theta$, and $\Psi$, respectively, we can exploit this observation performing the following simple steps:


\begin{enumerate}
  \item Employing $\left.\Psi\right|_{\partial\Omega}=0$ as a Dirichlet boundary condition, solve a preliminary Poisson equation for $\Psi(\mathbf{x})$ in the bulk volume, $\mathbf{x}\in\Omega$, using the standard MG method.
  \item Obtain the surface ``screening'' mass density, $\sigma(\partial\Omega)$, as $(4\pi G)^{-1}\nabla^2\!\left.\Psi\right|_{\partial\Omega}$, ~(assuming $\Psi=0$ in the exterior).
  \item Next, evaluate the surface potential, $\left.\Theta\right|_{\partial\Omega}$, via a convolution integral (on the surface, $\partial\Omega$) with a suitable Green's function, using $\sigma(\partial\Omega)$ as the source term.
  \item Applying $\left.\Phi\right|_{\partial\Omega} = -\left.\Theta \right|_{\partial\Omega}$ ~as a boundary condition, solve for the actual interior gravitational potential, $\Phi(\Omega)$, using a second invocation of the standard MG method.
  \end{enumerate}


As already pointed out by \citet{2019ApJS..241...24M}, special care has to be taken with respect to the third step, where the Green's function needs to be chosen such that it is an exact inverse with respect to the {\it finite-difference} discretisation of the Laplace operator in the discrete Poisson equation (used in the second step). We hence term this desired Green's function the ``FDGF'', and in the following describe a means of obtaining it for a given spherical-polar mesh geometry. Far away (in practical terms, at least 16 grid points) from the source point, the FDGF will of course approach the classic free Green's function of the Laplace operator, which is given by the inverse modulus of the separation vector. For intermediate separations, we find it practical to moreover use another discrete Green's function, which we term ``finite-volume'' Green's function (FVGF). This one assumes a semi-analytic solution \citep[following][and derived in appendix~A]{2014CeMDA.118..299H} for the gravitational potential of a homogeneous mass distribution within a mesh segment (hence the chosen name). One might naively conjecture that the FVGF should provide a useful surface potential. We will, however, empirically see that the FVGF, compared with the FDGF, has a too shallow core, which makes it unsuitable to be used for step 3, above.


\begin{figure}
  \centerline{
    \includegraphics[width=\columnwidth]{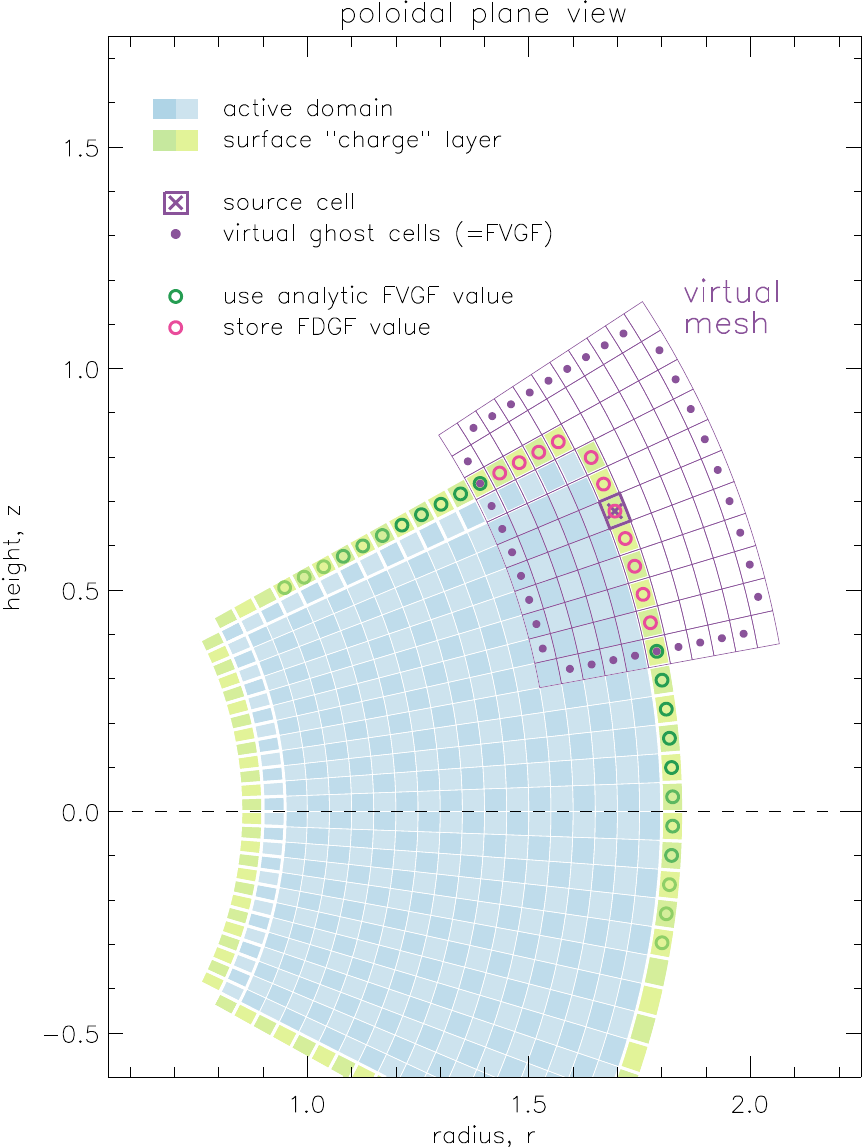}
  }
  \caption{{\it Illustration of the auxiliary spherical-polar mesh.} For each cell in the surface layer (light green), we create a virtual mesh (purple) centred at the cell of interest (marked `$\times$'). We then initialise all cells with their corresponding FVGF value. While the outermost layer of cells is used as a Dirichlet BC, we update interior points with a simple RB-SOR until convergence is reached. Depending on whether we are inside\,/\,outside of the virtual mesh, we finally assign FD\,/\,FV values (open circles) for the Green's function, respectively.}\label{fig:meshes}
\end{figure}


We illustrate the process of obtaining the FDGF in \Figure{meshes}, where we show a slice of the poloidal plane. While not shown here, the auxiliary mesh extends over the full {\it azimuthal} domain, where we moreover assume periodic boundary conditions.\footnote{The case of a p-fold azimuthal symmetry ---that is, for the case of a computational domain that only covers $2\pi/p$ in azimuth--- is a straightforward extension, discussed in \citet{2019ApJS..241...24M}.}

The initial step of obtaining the FDGF is only required at the very start of the simulation, and the obtained FDGF could in principle even be stored on disk for future use with identical mesh geometry. It is a scalar function of three independent variables, ~i) the linearly indexed coordinate along the poloidal circumference (made up of the inner/outer radial, as well as, the lower/upper latitudinal surfaces) of the location where the screening potential, $\left.\Theta\right|_{\partial\Omega}$, is to be evaluated, ~ii) the corresponding index of the coordinate where the contribution of the source density, $\sigma(\partial\Omega)$, is augmented, and ~iii) the azimuthal grid separation, $k-k'$, of the two points.\footnote{This last dimension collapses the two individual azimuthal indices, $k$ and $k'$ owing to the translation symmetry of the GF in the toroidal coordinate, $\phi$.} For practical purposes, we in fact pre-calculate and store the {\it Fourier coefficients} of the FDGF, which can directly be used in the convolution integral step employing an FFT (see the next subsection for details).

As a free parameter of our method, we can choose the number of cells that are added to each side of the source point when creating the auxiliary `virtual mesh'. We call this parameter $n_{\rm pad}$, representing the `padding' dimension. On the perimeter of the auxiliary mesh, we use the FVGF potential value as a Dirichlet boundary condition. Similarly, we use it as the initial condition on the entire domain. The latter is slightly preferable compared with the continuous Greens' function, which has a divergence at the reference point. Evaluating the FVGF is potentially expensive, but we empirically find that the integral (\ref{A.eq.3}) ---see the appendix--- only requires very few quadrature points to be evaluated at sufficient accuracy in this context.

\citet{2019ApJS..241...24M} discuss potential issues when the auxiliary grid intersects the polar axis. Amongst other things, they propose {\it ad hoc} to apply von Neumann boundary conditions at the inner radial boundary. We generally find the concern of the padding interfering with the central axis to be ameliorated compared to their approach. This is because we use the FVGF for the periphery values and simply turn any cells that interfere with the axis to be passive, that is, leaving them at their FVGF initial value. This means that we simply treat them as Dirichlet boundary values and accordingly do not update them in the solver iteration used for obtaining the FDGF. Moreover, we find that decent convergence is often obtained at $n_{\rm pad}\simeq 10$, as compared to the customary sixteen grid cells \citep{1977JCoPh..25...71J} that are empirically found to be required for asymptotically matching the continuous Green's function.

To obtain the FDGF on the auxiliary mesh, we implement a simple red-black (RB) successive over-relaxation to solve the Poisson equation. This may be seen as rather computationally inexpedient. Luckily, the discrete Green's function can be stored in memory for the duration of the time evolution sequence -- rendering the initial computational overhead at the start of the simulation insignificant. Moreover, the initial overhead is vastly outweighed by the simplicity of the screening-mass approach, where the complexity of the convolution integral \citep[compared, e.g., with][]{1999ApJ...527...86C} is reduced from a volume integral to a surface integral. In terms of the real-world computational scaling, this seemingly subtle difference is considerable.


\begin{figure}
  \centerline{
    \includegraphics[width=\columnwidth]{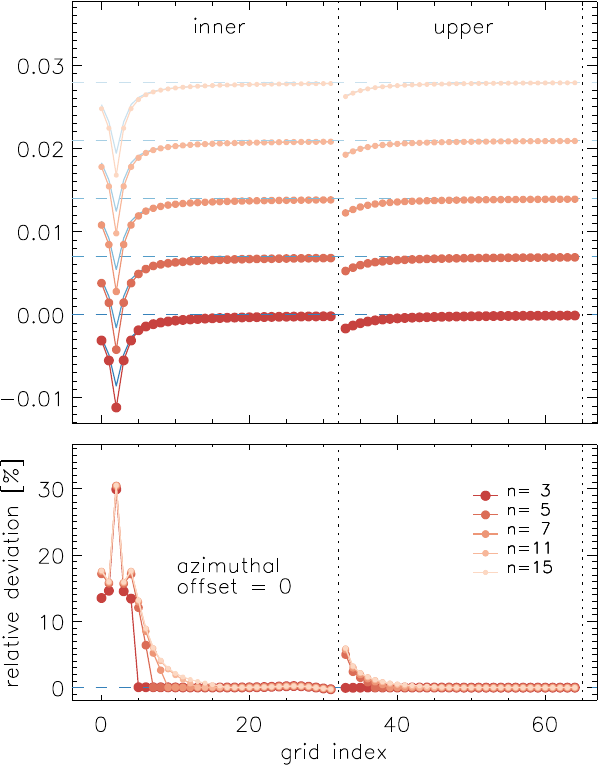}
  }
  \caption{{\it Shape of the FD discrete Green's function.}
    We show examples of the finite-difference discrete GF (red dots), obtained with mesh padding of $n$ cells, in comparison to the FV kernel (blue solid lines) obtained via the semi-analytic mesh-segment solution. Both panels are for a source point three grid cells down from the inner/upper corner. Curves are offset for clarity; the respective zero level is indicated by a dashed line for each curve. Relative deviations (up to 30\% at the location of the source) are substantial at $|k-k'|=0$.}\label{fig:fdgfzero}
\end{figure}


We now aim to illustrate the qualitative difference between the FVGF, on the one hand, and the FDGF, on the other hand. This is done in Figure~\ref{fig:fdgfzero}, where the left panel shows the GF for the inner radial (i.e., $r={\rm const.}=r_{\rm min}$) boundary, whereas the right panel illustrates the adjacent upper vertical (i.e., $\theta={\rm const.}=\theta_{\rm min}$) boundary. For this representative example, we chose a typical domain for a simulation representing the inner region of a protoplanetary disk, that is, we have a radial range of $r \in [2,8]$, spanning $\theta = 90^\circ \mp 20^\circ$ in co-latitude, and $\phi = 0$-$360^\circ$ in azimuth, at a fiducial resolution of 32$\,\times\,$32$\,\times\,$128.

In Fig.~\ref{fig:fdgfzero}, the ordinate spans the discrete grid location (as indexed in the linear vector along the circumference) of the point at which the potential is to be evaluated. For clarity, the abscissa has been offset to separate the various curves for varying $n_{\rm pad}$. At zero separation ---i.e., where the continuous GF would diverge--- we find a 30\% deviation between the FVGF and the FDGF, illustrating that the latter has a substantially more pronounced cusp, reflecting the compact stencil of the discrete Laplacian used to obtain the FDGF. In comparison, the homogeneous mass distribution inside a single cell (assumed for the FVGF) leads to a more diffuse kernel. As can be seen in the lower panel of Fig.~\ref{fig:fdgfzero}, the overall shape of the FDGF converges for $n_{\rm pad} \simeq 10$.


\begin{figure}
  \centerline{
    \includegraphics[width=\columnwidth]{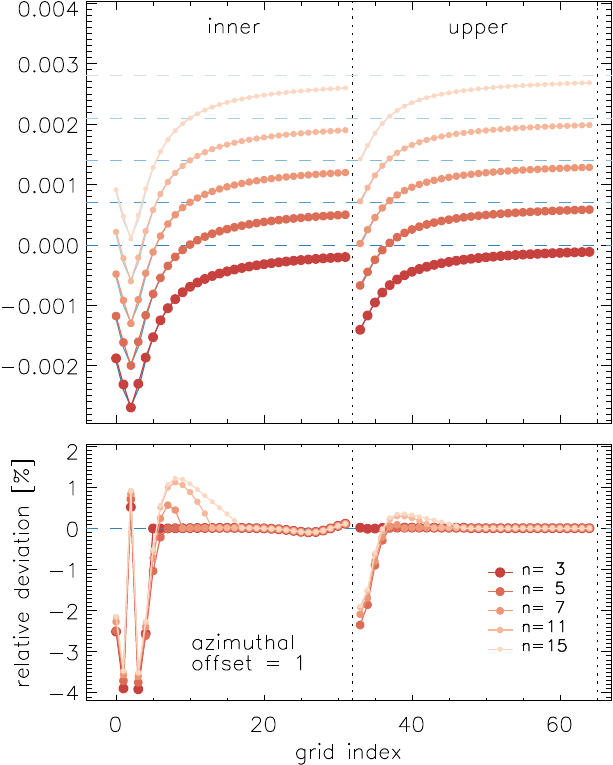}
  }
  \caption{{\it Shape of the FD discrete Green's function.} Same as \Fig{fdgfzero}, but for an azimuthal offset, $|k-k'|=1$. Here, relative deviations from the FV Green's function are at the percent level, and notably do not fully converge with $n$.}\label{fig:fdgfone}
\end{figure}


In Figure~\ref{fig:fdgfone}, we show the same comparison, but now at an azimuthal offset of one grid cell. The deviations from the FVGF are now at a level of less than 5\%. On the other hand, the results do not appear to be fully converged with respect to $n_{\rm pad}$, which however does not seem to have an influence on the usefulness of the FDGF within the convolution integral, as demonstrated by our convergence study in the next section.


\subsection{Implementation details}


We start with a practical note regarding the convergence of the MG solver in steps 1 and 4 (of the procedure outlined above), that is, for the case of time-dependent simulations. While, in terms of the computation, both operations are performed on the same array in memory, we found it advantageous to cache the respective solutions $\Psi^{(n-1)}(\Omega)$ and $\Phi^{(n-1)}(\Omega)$ from the previous timestep, $t^{(n-1)}$, separately. We then initialise the MG iteration cycle from the {\it matching} previous solution, rather than from the last available one. This generally accelerates the convergence of the MG solver considerably.

The crucial reduction in complexity of the described surface-screening approach lies in the convolution integral step, which (compared with the CCGF method) has a preferable scaling\footnote{That is, when exploiting a spectral decomposition plus the FFT algorithm.} $\approx \mathcal{O}(n^3)$. This is achieved because the surface density, $\sigma(\partial\Omega)$ occupies a two-dimensional region, compared with the entire volume density, $\rho(\Omega)$, used in CCGF.

In their description of step 3, \citet{2019ApJS..241...24M} introduce case distinctions regarding the inner/outer and lower/upper boundaries, which makes both the implementation and notation rather cumbersome (as one needs to consider all pairs of surfaces in the convolution integral). We avoid this complication by collecting the four individual boundaries into a linear vector. That way, the book keeping is reduced to a single linear grid index per argument in the GF. The major complication in the practical implementation stems from the requirement of distributed domains and MPI. Replicating the surface vectors as global variables is of course prohibitive in terms of the memory footprint. Therefore we introduce a partitioning along the vector index, and store the surface charge and potential, respectively, on the MPI rank that is native to the corresponding grid location.

The algorithm for calculating the surface potential can then be broken down into the following steps:

\begin{enumerate}
  \item Evaluate the surface screening mass, $\sigma$, and store it into the partitioned, linearly indexed vector, $\sigma(p,i',k)$, where $p$ denotes the partition index, $i'$ the ($p$-local) poloidal index, and $k$ the azimuthal grid index.
  \item $^\star\,$Calculate and store the discrete Green's function, $\mathcal{G}\, (p,i,i',k)$, where $i$ and $i'$ are indices along the surface (note that while $i$ is global, $i'$ is local to partition $p$).
  \item $^\star\,$Store the Fourier coefficients $\widehat{ \mathcal{G}} \,(p,i,i',m)$ of the discrete GF and release the memory used for $\mathcal{G}\,(p,i,i',k)$.
  \item For each partition, $p$, along the surface, evaluate the Fourier coefficients $\,\widehat{\sigma}(p,i',m)$ of the screening mass.
  \item For each point, $i$, along the entire surface, evaluate the surface potential $\Theta\,(i,k) \equiv {\rm FFT}^{-1}\,\big[\widehat{\Theta} \,(i,m)\big]$ via the GF convolution integral -- represented by a multiplication of the respective Fourier coefficients, that is, $\widehat{\Theta}\,(i,m) = \widehat{\mathcal{G}}\,(p,i,i',m)\;\times\;\widehat{\sigma}\,(p,i',m)$ -- note that this only adds the contribution from the density of the local partition, $p$.
  \item Augment the contributions to $\Theta\,(i,k)$ from all partitions, exploiting the superposition principle.
\end{enumerate}
Note that the steps 2 and 3, marked by an asterisk ($^\star$), are only executed once at the beginning.


\begin{figure*}
  \centerline{
    \includegraphics[width=0.7\textwidth]{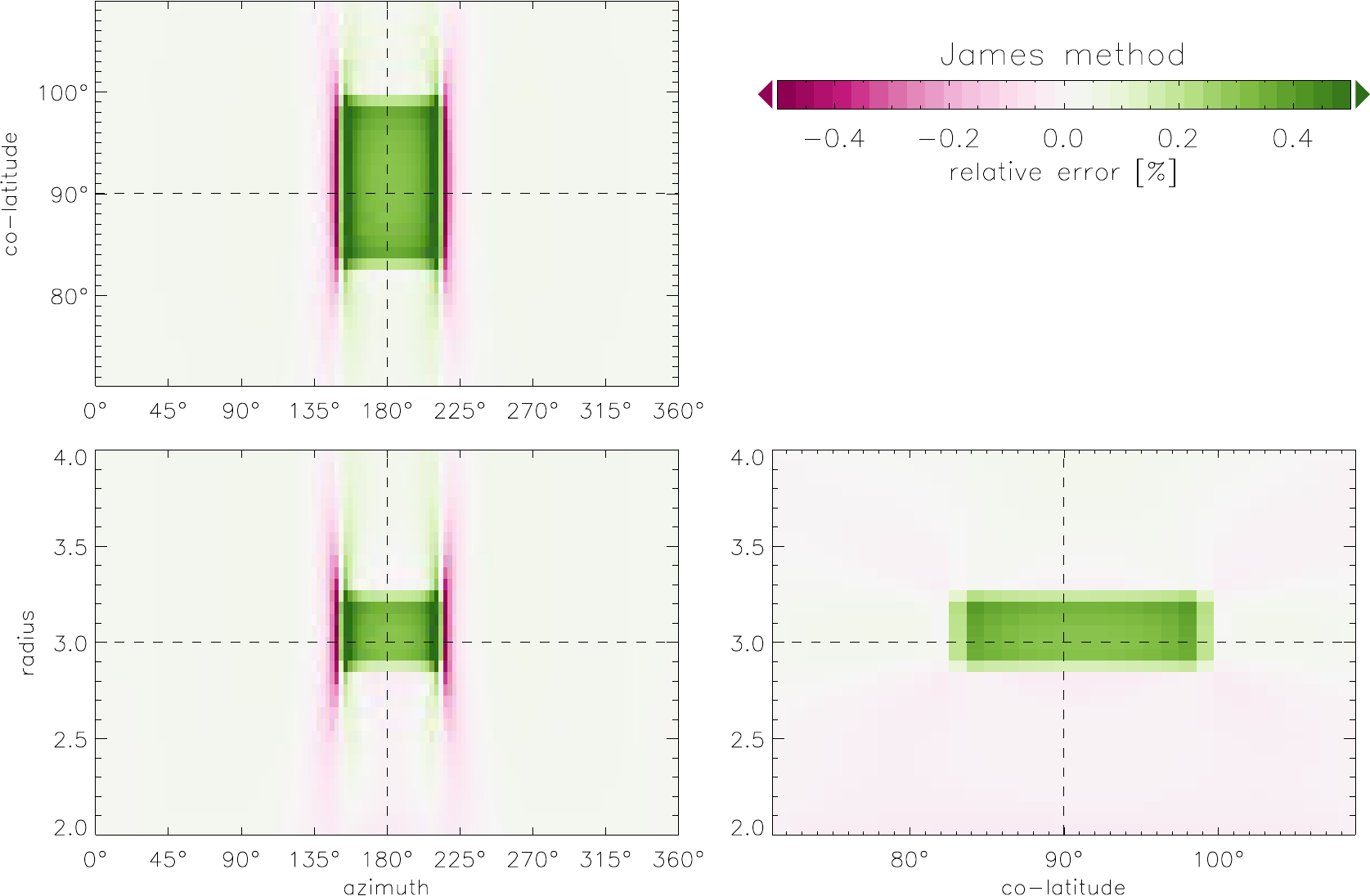}
  }
  \caption{{\it Relative error for a mesh-segment test.} We show the figure for the James method as a representative example; results for the CCGF method are essentially indistinguishable.}\label{fig:dev_James}
\end{figure*}


In our current implementation of the GF convolution integral, we use the function \texttt{fftw\_mpi\_plan\_many\_dft()} of the \textsc{fftw3} library \citep[see][]{2005IEEEP..93..216F}, for one-dimensional FFTs along the azimuthal direction. Specifically, we make use of the \texttt{FFTW\_MPI\_SCRAMBLED\_\{IN|OUT\}} flags to reduce MPI overheads. In this case, the Fourier coefficients are stored in a ``scrambled'' order, internally. This does however not affect the convolution operation, since both sets of Fourier coefficients are affected in the same way. Despite this optimisation, there are substantial intrinsic MPI overheads in one-dimensional FFTs as the algorithm, of course, involves all elements of the data. Yet, the MPI version of the FFT still outperforms our initial implementation using \texttt{MPI\_Allgather()} followed by an MPI-serial FFT.
For the case of uniform meshes, we make use of a columnar MPI communicator along azimuth, using the \texttt{MPI\_Comm\_split()} function, as well as a dedicated surface communicator, that is created by selecting the affected MPI ranks via \texttt{MPI\_Group\_incl()} and then using \texttt{MPI\_Comm\_create()}. We are aware that this is not optimal in terms of load balancing, as the interior MPI ranks in the 3D block decomposition are idle during the computation of the surface potential. The final augmentation of the local gravitational potential is done using the standard \texttt{MPI\_Allreduce()} function, which is immensely straightforward and implicitly exploits the library optimisations for collective communications.

It is not obvious whether a considerably more involved scheme with improved work sharing is practical or even feasible in view of the additional overheads from the then required memory transpositions. In practical terms, the expediency of the scheme can be judged by the benchmarking results presented in Section~\ref{sec:bench}.


\section{Convergence study} \label{sec:conv}


To verify our implementation, we use a static test problem that features a step-wise mass distribution with a constant density $\rho_\lhd=1$ inside a spherical-polar mesh segment (or sector), located at a fiducial location $\mathbf{r}=(r_\lhd,\theta_\lhd,\phi_\lhd)$ in the grid interior. By default, \nirv uses SI units, but code units can be specified. In that case, the numerical value of the gravity constant, $G$, in SI units is divided by a factor $\hat{l}^3/(\hat{m}\,\hat{t}^2)$, where $\hat{l}$, $\hat{m}$, and $\hat{t}$ are unit conversion factors (to SI units) for length, mass and time, respectively. For protoplanetary systems we typically use $\hat{l}=1\,{\rm au}$,~ $\hat{m}=1\,{\rm M_\odot}$,~ and $\hat{t}=1\,{\rm yr}/2\pi$.

The sector is conveniently given by coordinates $(r,\theta,\phi)$ enclosed in the intervals $r \in [r_1,r_2]$ in radius, $\theta \in [\theta_1, \theta_2]$ in latitude, and $\phi \in [\phi_1, \phi_2]$ in azimuth. We arbitrarily chose the six constants $r_1$, $r_2$ as $r_\lhd-1\Delta_0 r$,~ $r_\lhd+2\Delta_0 r$, moreover $\theta_1$, $\theta_2$ as $\theta_\lhd-3 \Delta_0\theta$,~ $\theta_\lhd+4\Delta_0\theta$, and $\phi_1$, $\phi_2$ as $\phi_\lhd-5\Delta_0\phi$,~ $\phi_\lhd+6\Delta_0\phi$, respectively. The above choice of numbers ensures that any sort of accidental grid symmetry is explicitly excluded. Choosing a mesh sector based on the {\it coarsest} grid resolution $\Delta_0 r$, $\Delta_0\theta$, and $\Delta_0\phi$ in each coordinate direction, respectively, has the benefit of being able to represent the mass density distribution in an identical fashion on all progressively finer grids, that is, when performing the convergence study by doubling the grid resolution. This entirely avoids the issue that any aliasing of the mass distribution on the computational grid (as, e.g., in the case of a solid sphere, for which a simple analytic solution exists) may blur the resolution-dependence of the obtained numerical error.


\begin{figure}
  \center{
    \includegraphics[width=\columnwidth]{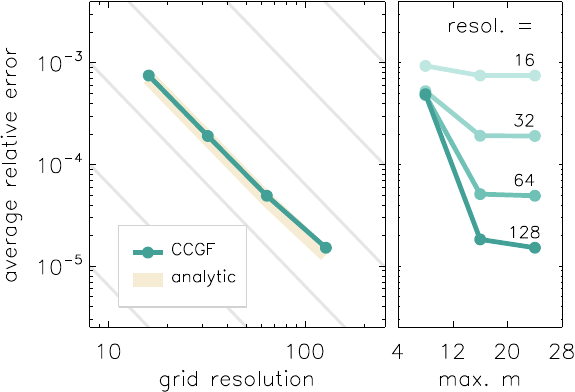}\\[14pt]
    \includegraphics[width=\columnwidth]{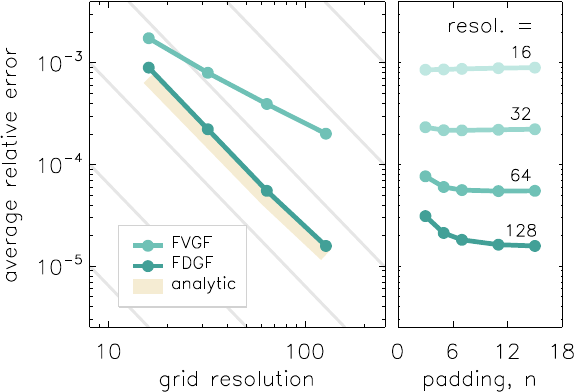}\\[-1pt]
  }
  \caption{{\it Convergence study.} We show the mean relative error as a function of grid resolution for the CCGF method (top left), as well as for the James method (bottom left panel) -- the latter using the FV-\,/\,FDGF, respectively. The key parameter dependencies are shown on the right-hand side.}\label{fig:conv}
\end{figure}


As a reference solution, we use our own spherical-polar extension ---see Eqn.~(\ref{A.eq.3})--- of the semi-analytic expression for a {\it cylindrical} mesh segment given originally by \citet{2014CeMDA.118..299H}. For reference, we document the functional form of this novel solution in the appendix. The semi-analytic solution evaluates $\Phi(\mathbf{r})$ by means of the given expression. In general, 32 quadrature points and using a simple Romberg integrator achieves a sufficient accuracy for our purposes. If available, the GNU scientific library \citep{gough2009gnu} is used for evaluating the appearing elliptic integrals -- otherwise a simple AGM algorithm can be used.

For comparison, we perform the convergence test both for the CCGF method (for $m_{\rm max}$ up to 24), as well as for the James method (with varying values $n_{\rm pad}$ for the padding of the FDGF). As an example, the relative deviation from the semi-analytic solution for a spherical-polar mesh segment is shown in \Figure{dev_James}, for the case of the James method at a fiducial resolution of 32$\,\times\,$32$\,\times\,$128 mesh points. The error distribution illustrates that the peak of the error is directly adjacent to the discontinuity in density at the edge of the mesh segment (affecting its interior), and errors near the computational boundary are only moderate in comparison.

We present the results of our convergence study in \Figure{conv}, where the CCGF method is shown in the top panels, and the James method is shown in the bottom panels. The grid resolution corresponds to $N_r=N_\theta$, while $N_\phi$ is four times that value, producing a roughly isotropic grid spacing at the centre of the domain. Except when using the FV Green's function (`FVGF', which presumably has a too shallow core), both methods produce near second-order accuracy. In both panels, we highlight the optimal scaling by a band labelled `analytic', for which we have simply used the reference solution as a Dirichlet boundary condition. Focusing on the left-hand panels, we can see that both methods produce near-optimal convergence, when appropriate parameter values are chosen -- that is, $m_{\rm max}=24$ in the case of the CCGF method, and $n_{\rm pad}=15$ for the James method with FDGF kernel. Convergence is impeded, if inappropriate values are chosen. This is illustrated on the right-hand side, where we plot curves against the free parameter of each method and give the grid resolution by the label. Note that the rightmost set of points corresponds to the curve shown in the left-hand panel. For the simple test problem at hand (which essentially is $m=1$), the CCGF already requires $m_{\rm max}\ge 16$ for appropriate convergence (see the upper right panel).

In comparison, when using James' method, the FDGF is relatively insensitive to the extent of padding (lower right panel), that we apply around the core of the discrete Green's function. This is with the exception of very high resolution, where a dependence on the $n_{\rm pad}$ parameter becomes noticeable.


\section{Benchmarking} \label{sec:bench}


\begin{figure}
  \center{
    \includegraphics[width=0.95\columnwidth]{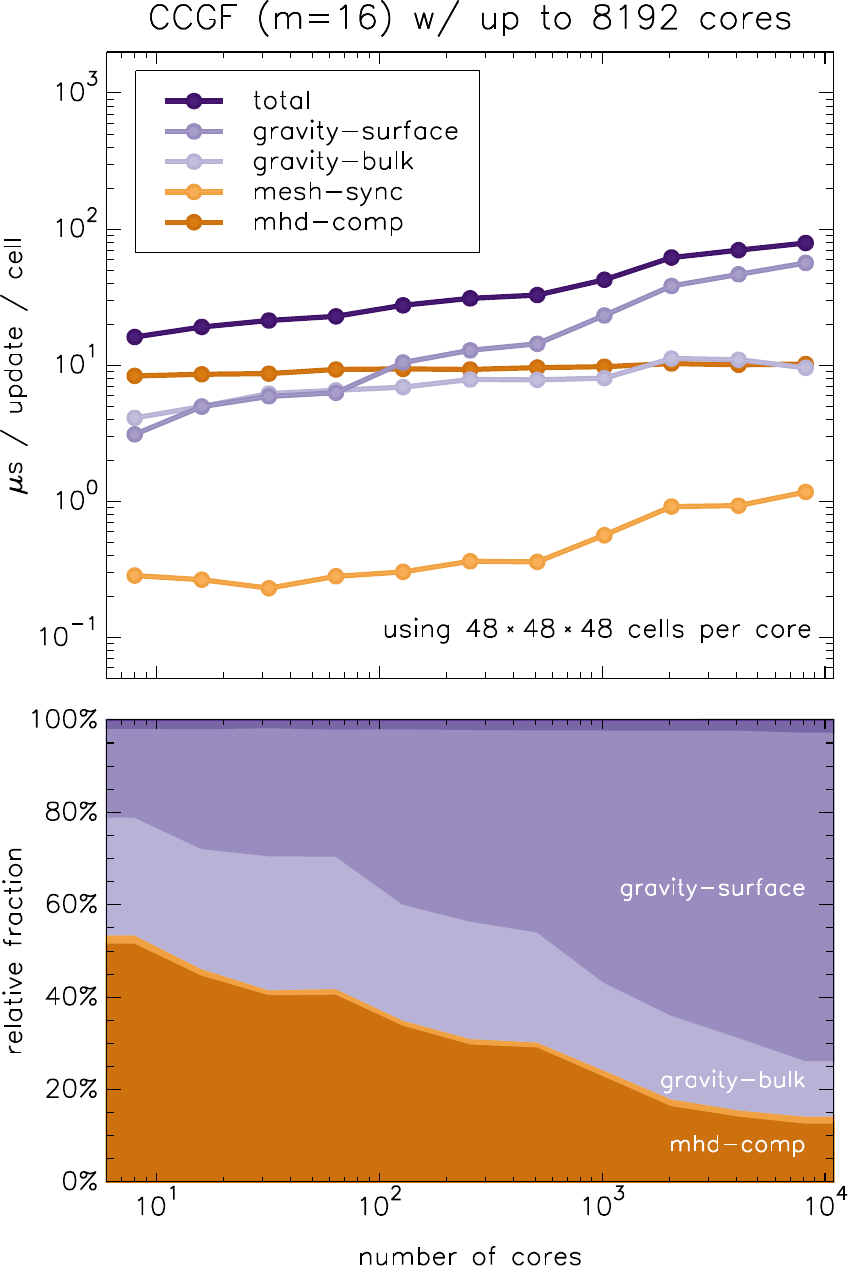}
  }
  \caption{{\it Weak scaling study (CCGF).} With a moderate workload of $48^3$ cells per MPI task, the MG solver consumes roughly an equal amount of time than the MHD solver. The $\mathcal{O}(n^4)$ scaling of the CCGF method becomes noticeable as a linear increase in the update time per cell -- that is, after dividing by the number of grid cells $\sim\mathcal{O}(n^3)$. }\label{fig:mpi_scaling_ccgf}
\end{figure}


We have benchmarked our implementation on the HLRN facility ``Emmy'' which features two 48-core Intel Xeon Platinum 9242 per compute node.\footnote{Note, however, that we have used $\le\!64$ cores per node for benchmarking.} Specifically, \nirv was built against version 19 of the Intel MPI library, using version 3.3.8 of the FFTW3 library.

In order to minimise the potential for spurious trends in the scaling test, we have kept the number of MG solver iterations (including the number of smoothing steps on the base mesh) fixed when running a benchmark simulation. We use a representative value of eight V-cycle iterations per MG update, and note that for typical applications this is probably still a somewhat conservative estimate.

In \Figure{mpi_scaling_ccgf}, we showcase the weak-scaling behaviour of our reference implementation of the CCGF method. For increased problem size, the leading-order $\mathcal{O}(n^4)$ scaling reveals itself as a linear growth in the time required for the surface gravity computation. This makes the use of the method doubtful when reaching a core count of a few thousand. And that is at the rather moderate angular accuracy of $m_{\rm max}=16$, which may not at all be suitable for simulations that develop high-wavenumber azimuthal structures.


\begin{figure}
  \center{
    \includegraphics[width=0.95\columnwidth]{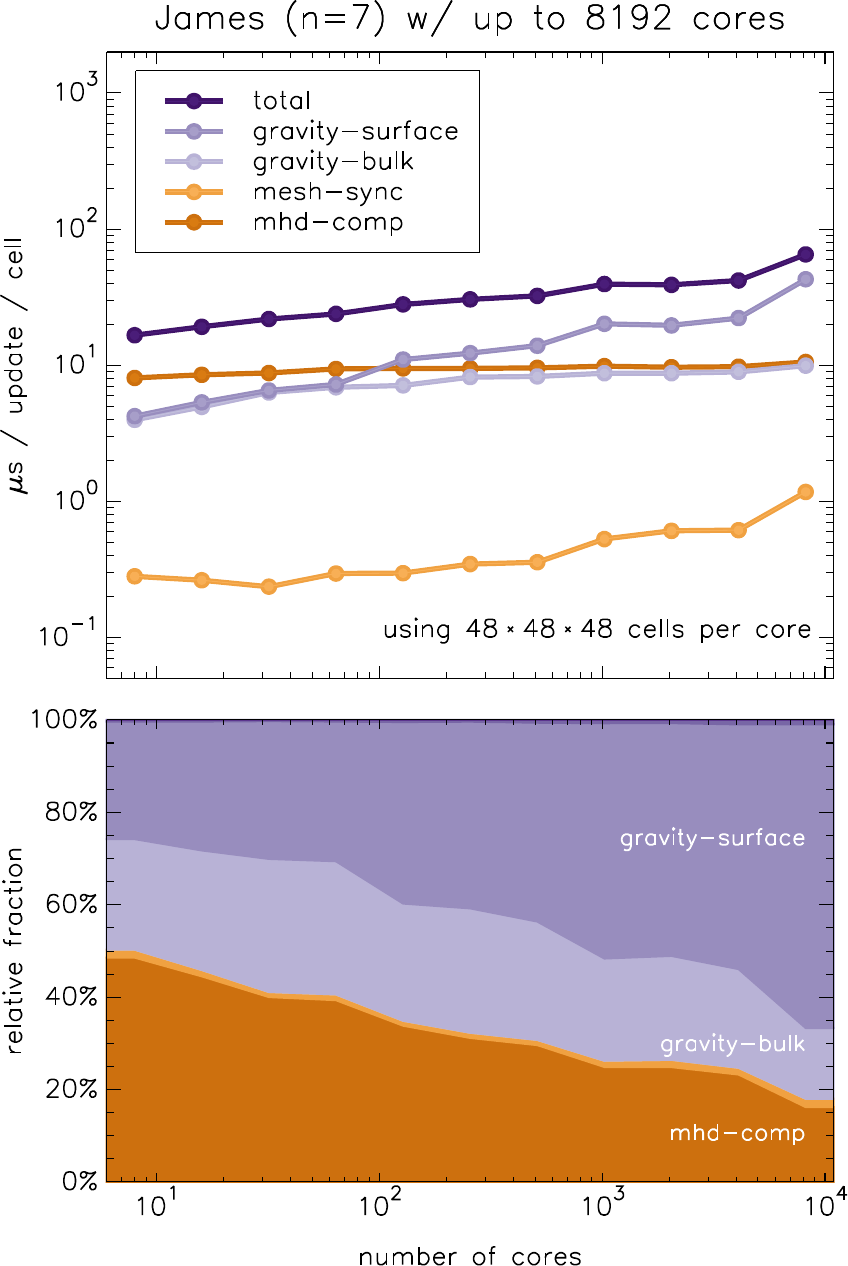}
  }
  \caption{{\it Weak scaling study (James-FDGF).} Overheads from the convolution operation for the surface gravitational potential become noticeable at $>\!64$ cores (the node boundary) and remains tolerable up to a few thousands of cores.}\label{fig:mpi_scaling_james}
\end{figure}


In \Figure{mpi_scaling_james}, we demonstrate the weak-scaling behaviour of our implementation of the surface-density method using a FDGF discrete Green's function with padding parameter\footnote{We note that, unlike for the CCGF method, the parameter $n_{\rm pad}$ does not have any influence on the run-time performance, but only on the initialisation time, which we however exclude from the benchmarking results.} $n_{\rm pad}=7$. The scaling test was performed at a fixed workload of $48^3$ cells per MPI task, which we deem a typical real-world use case. We overall regard the observed total update time of $\simeq 20-40\mu s$ per cell as satisfactory, given that it can be maintained up to a core count of a few thousand.

As it is based entirely on non-blocking MPI point-to-point communications, the MG Poisson solver introduced in \citet{2005A&A...435..385Z} demonstrates an excellent MPI scaling behaviour up to the number of cores tested here (see the curve labelled `gravity-bulk'). For the chosen workload, we obtain an excellent performance of $10\mu s$ per update for the Poisson solver alone. The computationally expensive procedure of obtaining the gravitational potential on the poloidal domain boundaries (labelled `gravity-surface') already becomes noticeable when using hundreds of MPI tasks. Until then, the operations related to the convolution integral are minimal, and the curve essentially follows the one of the MG solver (which comprises the initial step within the compound algorithm). We notice a marked step-like behaviour after each 8-fold increase in the core count. This might be related to the way that we double the resolution along $r$, $\theta$, and $\phi$ in a round-robin fashion. Another explanation might be the node boundaries on the compute cluster. This would indicate the noticeable role of MPI communications in the algorithm.


\begin{table}
    \def\arraystretch{1.125}
  \setlength\tabcolsep{3pt}
  \begin{tabular}{p{2.5cm} >{\bf}ccc >{\bf}ccc >{\bf}cc}
    \hline
                   &  \multicolumn{2}{c}{1024}  &
                   &  \multicolumn{2}{c}{2048}  &
                   &  \multicolumn{2}{c}{4096}  \\ \hline \\[-10pt]
            CCGF (m=16)  &   23.2  &   42.5  &
                   &   38.2  &   61.9  &
                   &   46.6  &   70.1  \\
            James (n=7)  &   20.2  &   39.5  &
                   &   19.6  &   39.0  &
                   &   22.2  &   42.0  \\
    \hline
  \end{tabular}
    \caption{{\it Weak scaling study comparison.} We here contrast the key timings (in $\mu$s per cycle) of the two methods. For each core count, we show the value for `gravity-surface' (left column, boldface) and the `total' (right column).}
  \label{tab:bench}
  \end{table}


In Table~\ref{tab:bench}, we compare the two methods at core counts of 1024, 2048, and 4096 MPI tasks, respectively. Whereas the two methods perform equally well at around a thousand cores, the CCGF algorithm quickly loses ground beyond that, owing to its unfavourable scaling behaviour. At around four thousand cores, CCGF already uses more than twice as much time compared with the James method.

In conclusion, we may speculate that the most prominent aspects that limit a better overall scaling (of both methods) are twofold: firstly, the one-dimensional FFT along azimuth operates on distributed memory, which amounts to a noticeable MPI overhead. Secondly, the computed surface potential requires a substantial MPI reduction, that is, across all MPI ranks touching the poloidal surface (and sharing a given partition in azimuth). While we currently use a collective MPI operation for this, it is unclear whether an improved performance can be obtained by point-to-point operations. That having said, its better scaling behaviour clearly favours the James method over the CCGF approach.


\section{Conclusions} \label{sec:conclusions}


In this article, we have presented our implementations of two different computational approaches for obtaining the surface gravitational potential on spherical-polar meshes -- specifically treating the inner radial boundary that cannot be approximated with a standard multipole expansion of the potential.

The first method, termed CCGF \citep{1999ApJ...527...86C} is fairly straightforward to implement but relies on a volume integral over the gas density to evaluate the convolution operation of the Green's function approach. Its computational expense can be curbed by arbitrarily introducing a truncation in the azimuthal wavenumber, but the merit of this remains questionable in terms of the unknown level of degradation in the scheme's accuracy, depending on the application at hand.

The second method, is based on a surface screening mass \citep{1977JCoPh..25...71J} and conceptionally a little more involved. We here present a reference implementation for uniform meshes and test it in the realm of distributed-memory parallelism. Care has to be taken in obtaining a suitable discrete representation of the Green's function associated with the Laplace operator in the Poisson equation.

If parameters are chosen adequately, both methods show essentially second-order convergence for the simple test problem of a mesh sector with uniform mass density \citep[see][as well as our appendix, below]{2014CeMDA.118..299H}. In terms of the scaling with the number of cores (in a weak scaling study with fixed number of solver iterations), the latter method clearly shows an edge compared to the former. This is because the convolution integral has one fewer dimensions, which makes a noticeable difference for increased grid counts. Finally, James' method does not rely on any sorts of trade-offs in terms of its level of approximation, as does CCGF, where one typically truncates the azimuthal wavenumber for improved computational expediency. The only minor limitation of the screening mass method is that obtaining the discrete Green's function may become difficult when the (auxiliary) grid is extending close to the coordinate axis of the polar mesh. If possible, this case should be avoided, not least because the grid anisotropy does also affect the convergence rate of the MG solver (which currently uses a simple point smoother). Nonetheless, a small inner radius can also become problematic for the CCGF method, where the evaluation of the elliptic integral functions has to be done at sufficient numerical accuracy to avoid divergences. Overall, our analysis clearly favours the screening mass approach.


\section*{Appendix:}

\subsection{Gravitational potential of a homogeneous spherical mesh segment}
\label{sec:appendix}

\def\dd{{\rm d}}
\def\rp{r^{\prime}}
\def\thetap{{\theta}^{\prime}}
\def\phip{{\phi}^{\prime}}

We derive an expression for the gravitational potential of a spherical mesh segment given by $[r_1,r_2]\times[\theta_1,\theta_2]\times[\phi_1,\phi_2]$ with homogeneous density, $\rho$, in the interior.

\begin{figure}
  \center\includegraphics[width=0.75\columnwidth]{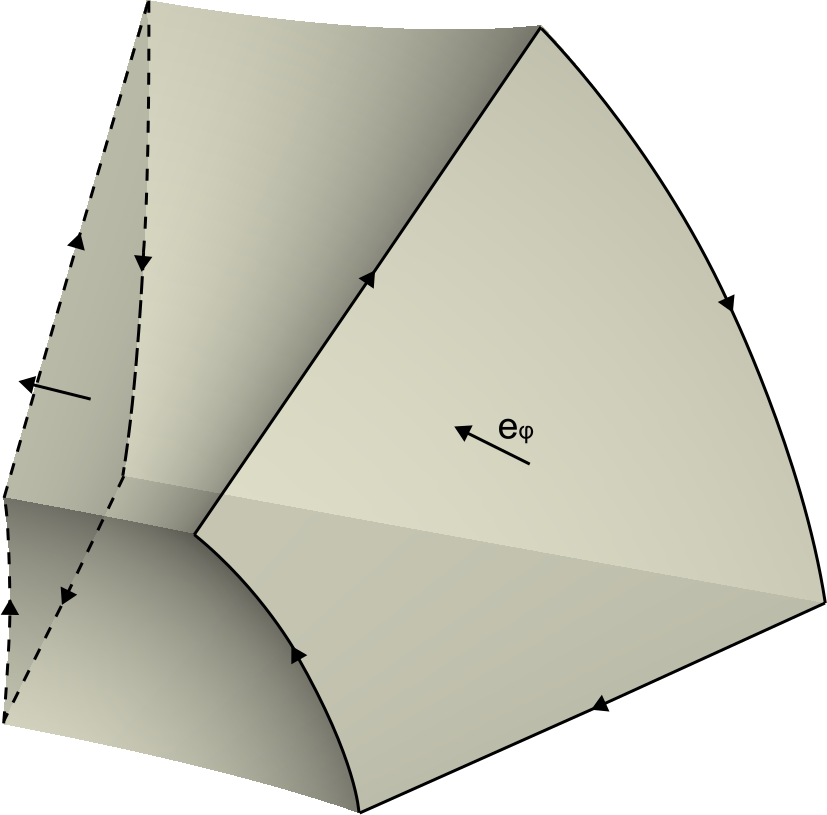}
  \caption{{\it Visualisation of the mesh segment,} and definition of the sense of orientation of the poloidal surfaces.}
  \label{fig:A.fig.1}
\end{figure}

Our motivation to do so is twofold. First, the presented segment solution is an ideal benchmark for a gravity solver in spherical coordinates since it can be computed with high accuracy at low computational costs to provide a reference with exact mass representation on the grid. To our knowledge, such a derivation has not been presented in the literature so far.  Second, when applied to a single mesh cell the segment solution reflects a finite-volume, non-singular analogue to the continuous GF and, therefore, better approximates the FDGF of Section~\ref{sec:screen} in the near-source limit than the continuous Green's function.

Our derivation closely follows the work of \citet{2014CeMDA.118..299H}, who derived an elegant expression for the gravitational potential of a homogeneous {\it cylindrical} mesh segment. We begin by expressing the inverse of the spatial separation, $|\mathbf{r}-\mathbf{r}^{\prime}|$, in spherical polar coordinates $(r,\theta,\phi)$, i.e.,
\smallskip
\begin{equation}
\frac{1}{|\mathbf{r}-\mathbf{r}^{\prime}|}=
\frac{k}{2}\frac{1}{\sqrt{r\sin\theta\;\rp\sin\thetap}}
\frac{1}{\sqrt{1-k^2\cos^2\!\big(\frac{1}{2}(\phi-\phip)\big)}}\,,
\end{equation}
where we define the parameter $k$ as
\begin{equation}
  k=\frac{2\,\sqrt{r\sin\theta\;\rp\sin\thetap}}         {\sqrt{r^2+{\rp}^2 - 2\,r\,\rp\cos(\theta+\thetap)}}\,.
\end{equation}
With this definition, the gravitational potential of the segment then reads
\begin{eqnarray}
  \Phi = & - G\rho\, & \int_{r_1}^{r_2}
              \dd\rp\, \int_{\theta_1}^{\theta_2}
              \frac{\dd\thetap\, {\rp}^2\!\sin\thetap\frac{k}{2}}                   {\sqrt{r\,\rp\sin\theta\sin\thetap}}
                   \nonumber\\
         & & \quad \int_{\phi_1}^{\phi_2}
                   \frac{\dd\phip}                   {\sqrt{1-k^2\cos^2\!\big(\frac{1}{2}(\phi-\phip)\big)}}\,.
\label{A.eq.1}
\end{eqnarray}
By introducing the substitution $2\beta^{\prime}=\pi-(\phip-\phi)$,
the $\phi$-integral yields
\begin{eqnarray}
  \int_{\phi_1}^{\phi_2}
  \frac{\dd\phip}{\sqrt{1-k^2\cos^2\!\big(\frac{1}{2}(\phi-\phip)\big)}} & = &
  \\ \nonumber
  -2\int_{\beta_1}^{\beta_2}
          \frac{\dd\beta^{\prime}}{\sqrt{1-k^2\sin^2\beta^{\prime}}}
          & = & 2\left(F(\beta_1,k)-F(\beta_2,k) \right)\,,
\end{eqnarray}
with $F(\beta,k)$ the incomplete elliptic integral of the first kind, defined by
\begin{equation}
  F(\beta,k)=\int_0^{\beta}
  \frac{\dd\beta^{\prime}}{\sqrt{1-k^2\sin^2\beta^{\prime}}}\,.
\end{equation}
It can moreover be shown that the potential, $\Phi$, can be expressed as the difference $\Phi=\Psi(\beta_1)-\Psi(\beta_2)$, with functions
\begin{equation}
  \Psi(\beta_\alpha)\ =\ -G\rho\int_{r_1}^{r_2}\dd\rp\int_{\theta_1}^{\theta_2}
  \dd\thetap \rp\sqrt{\frac{\rp\sin\thetap}{r\sin\theta}}k\;
  F(\beta_\alpha,k)\,.
\end{equation}
We can now apply Green's theorem to a $r\theta$-area, $S$, in the poloidal plane, with a normal vector $\mathbf{e}_{\phi}$. The theorem provides a relation between the surface integral of a curled vector function with a line integral of the function itself taken around its contour, that is,
\smallskip
\begin{equation}
  \iint_S
  \!\rp \dd\rp \dd\thetap \left[\frac{1}{\rp}
    \frac{\partial\,(\rp\!E_{\theta} )}{\partial \rp}
    -\frac{1}{\rp}\frac{\partial E_r}{\partial\thetap} \right]
  =\oint_{\partial S}\left(E_r \dd\rp +E_{\theta}\rp \dd\thetap \right)
\end{equation}
for a vector $\mathbf{E}=E_r\mathbf{e}_r+E_{\theta}\mathbf{e}_{\theta}$, and where the contour integral along $S$ has to be performed right-handed with respect to $\mathbf{e}_{\phi}$ (see \Fig{A.fig.1}).
Provided that functions $E_r$, and $E_{\theta}$ can be found such that
\begin{equation}
  \frac{1}{\rp}\frac{\partial(\rp E_{\theta})}{\partial \rp}
  -\frac{1}{\rp}\frac{\partial E_r}{\partial\thetap}=
  \sqrt{\frac{\rp\sin\thetap}{r\sin\theta}}\,k\,F(\beta,k)\,,
  \label{A.eq.2}
\end{equation}
then the potential can be represented simply as the difference of contour integrals at the upper and lower poloidal interface, respectively. That is,
\begin{eqnarray}
\Phi(\mathbf{r}) & = - G\rho\ \Big[\Big. &
  \oint_{\partial S_1}\left(E_r^{(1)} \dd\rp +E_{\theta}^{(1)}\rp \dd\thetap \right)
  \nonumber\\
  &&
  -\oint_{\partial S_2}\left(E_r^{(2)} \dd\rp +E_{\theta}^{(2)}\rp \dd\thetap \right)
  \ \Big.\Big]
\label{A.eq.3}
\end{eqnarray}
where $E_{r/\theta}^{(\alpha)} = E_{r/\theta} (r,\rp, \theta,\thetap, \beta_\alpha)$ are evaluated for $\beta_\alpha=\beta_1$, and $\beta_2$, respectively. Next, in order to find appropriate functions $E_r$ and $E_{\theta}$, we make the plausible ansatz
\begin{eqnarray}
E_r & = & E_R\sin\thetap + E_z\cos\thetap \nonumber \\
E_{\theta} & = & E_R\cos\thetap - E_z\sin\thetap\,.
\end{eqnarray}
where $E_z, E_R$ are the corresponding functions for the cylindrical segment
case as given by the expressions (27) and (28) in \citet{2014CeMDA.118..299H}. Note, however, that our convention introduces an overall minus sign, such that $E_z$ and $E_R$ are equivalent to $-M_0$ and $-N_0$, respectively.
Intuitively, our ansatz can be understood as a vector transform in the polar plane. The generalised expression of a spherical-polar sector can then be formulated as
\begin{eqnarray}
  E_r = & -\frac{1}{2}r\,\sqrt{\frac{\rp\sin\thetap}{r\sin\theta}}\,k
  & \Big[ \sin(\theta+\thetap)\;F \Big.
  \nonumber\\
  & & \Big. -\frac{2\sin\theta \cos\thetap}{k^2}\;(F-E)\ \Big]
  \nonumber\\
  & +\;\alpha\sin\thetap f & +\quad (1-\alpha) \cos\thetap g\,,
  \label{A.eq.4}
\end{eqnarray}
and
\begin{eqnarray}
  E_{\theta} = & -\frac{1}{2}r\,\sqrt{\frac{\rp\sin\thetap}{r\sin\theta}}\,k
  & \Big[\big(\cos(\theta+\thetap)-\frac{\rp}{r}\big)\;F\Big.
  \nonumber\\
  & & \Big. +\frac{2\sin\theta \sin\thetap}{k^2}\;(F-E)\ \Big]
  \nonumber\\
  & +\;\alpha\cos\thetap f & -\quad (1-\alpha)\sin\thetap g\,,
  \label{A.eq.5}
\end{eqnarray}
with $E(\beta,k)$ the incomplete elliptic integral of the second kind, and with additional functions
\begin{multline}
  f(r,\rp,\theta,\thetap,\beta) = \frac{1}{2}r\sin\theta\sin(2\beta) \\
  \operatorname{asinh}{\frac{r\cos\theta-\rp\cos\thetap}
    {\sqrt{(\rp\sin\thetap +r\sin\theta)^2-4\,r\sin\theta\;\rp\sin\thetap \sin^2\beta}}}\,,
\end{multline}
and
\begin{multline}
g(r,\rp,\theta,\thetap,\beta) = \frac{1}{2}r\sin\theta\sin(2\beta) \\
\operatorname{asinh}{\frac{\rp\sin\thetap +r\sin\theta\cos(2\beta)}
{\sqrt{(r\cos\theta -\rp\cos\thetap)^2+r^2\sin^2\theta \sin^2(2\beta)}}}\,,
\end{multline}
and with a free ``mixing'' parameter $\alpha\in [0,1]$.

By explicit computation of the lhs. of Eqn.~(\ref{A.eq.2}), it can be proven that Eqns.~(\ref{A.eq.4}), and (\ref{A.eq.5}) are indeed the desired functions. The algebra is simple but cumbersome and we dispense with it here. Note that $E_r$ and $E_{\theta}$ are regular functions compared to the original expression (\ref{A.eq.1}) for the potential containing a singularity. In the special case $\beta=\pi/2$ ---occurring on the $\phi$-areas bounding the segment--- the term $\ (\sin(\theta+\thetap) -2\sin\theta \cos\thetap/k^2) F(\beta,k)\ $ in Eqn.~(\ref{A.eq.4}) and the corresponding term $\ (\cos(\theta+\thetap) -\rp/r +(2\sin\theta \sin\thetap)/k^2)F(\beta,k)\ $ in Eqn.~(\ref{A.eq.5}) both go to zero in the limit $k\longrightarrow 1$ (i.e., when $r\longrightarrow \rp, \theta\longrightarrow \thetap$), despite the singularity $F(\pi/2,1)=\infty$.

Note, moreover, that by means of explicit numerical verification, the gravitational potential proves to be independent of the particular choice of the parameter $\alpha$ in functions $E_r$ and $E_{\theta}$. In practice, the expression (\ref{A.eq.3}) can be evaluated with high accuracy by application of any high-order quadrature rule to the contour integrals such as, for instance, a Romberg method.


\section*{Acknowledgements}


We thank the anonymous referee for a meticulous and useful report. Test computations were performed on AIP's \texttt{taurus} cluster node. The MPI scaling tests were done on machine \texttt{Emmy} at the HLRN centre in G{\"o}ttingen. This work was co-funded\,\footnote{Views and opinions expressed are however those of the author(s) only and do not necessarily reflect those of the European Union or the European Research Council. Neither the European Union nor the granting authority can be held responsible for them.} by the European Union (ERC-CoG, \textsc{Epoch-of-Taurus}, No. 101043302). The line plots in the article were created using the open-source GDL software package \citep{2022JOSS....7.4633P}. Any derived data produced from the simulations will be made available upon reasonable request.


\end{document}